David Christian, Mike Geelhoed & Nikolai Mokhov
July 13, 2011

# E906-SeaQuest MARS15 Simulation

The E906-SeaQuest spectrometer is designed to measure high energy muons produced in the forward direction by interactions of the 120 GeV Main Injector proton beam with a variety of targets. The spectrometer consists of two dipole magnets (both of which deflect charged particles in the horizontal plane) and a collection of tracking detectors. The first spectrometer magnet (FMAG) is a solid iron magnet. This magnet serves as a beam dump as well as a muon analysis magnet.

A series of MARS15 simulations were done by Nikolai Mokhov to verify and guide the design of steel and concrete shielding around FMAG and the target area immediately upstream of FMAG. The result of the last round of simulations is summarized here.

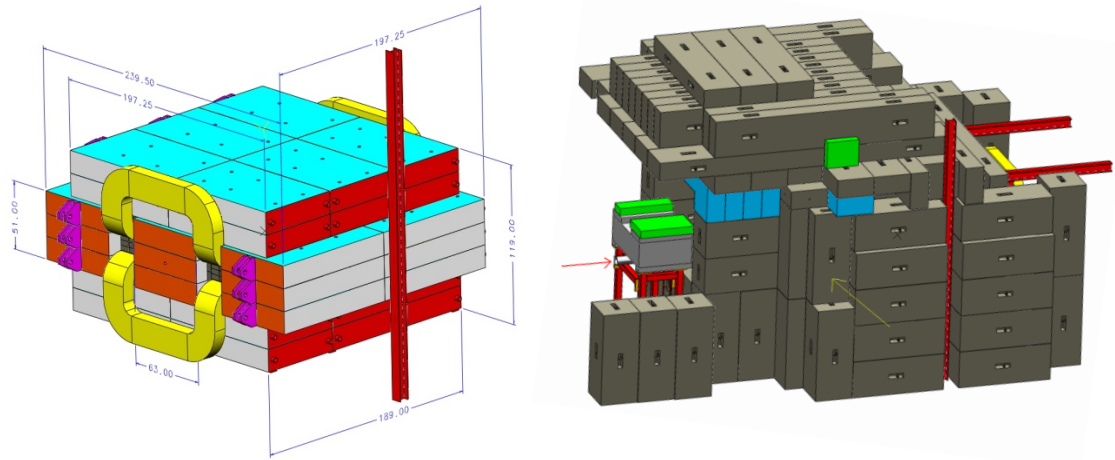

**Figure 1: Left) FMAG construction: the rectangular solid elements are soft magnet iron, the (yellow) coils are aluminum, and there are (grey) concrete bricks stacked between the coils. The red I-beam is part of the structure in NM4 that supports ductwork and cable trays. Right) FMAG shielding: the beam enters from the left as shown by the red arrow. A movable steel (gray) and concrete (green) shield is located upstream of the target cave. The targets and the magnet itself are shielded by a mixture of concrete and steel. Additional detail is shown in Figure 2.**

The MARS15 simulation used a model of FMAG and its surroundings. The model includes air gaps in the concrete shielding, the largest of which are required because of the geometry of the saddle coils. A small volume surrounding the beam line just upstream of the magnet is filled with borated polyethylene. The borated polyethylene extends into the air gap necessitated by the saddle coils. The top layer of blocks is intended to shield the roof and downstream end of NM4 in the event of a loss of beam accident well upstream of the target.

The MARS model of the solid iron magnet is less detailed than the model of the concrete shielding. There is a 2 inch diameter hole drilled 10 inches deep into the magnet iron at the upstream end. The beam dumps at the end of this hole. The aluminum coils are represented as simple rectangular solids. The area upstream of the magnet occupied by the coils is simply modeled as part of an air gap. Two more rectangular solids of appropriate volume represent the concrete bricks packed between the coils. Only the central 2 Tesla magnetic field is modeled. Since very few charged particles are transported outside the region of the vertical field, and essentially no charged particles other than muons escape the magnet iron, the magnetic return flux is not modeled. Representative pictures of the geometry of the MARS15 model are shown in Figure 2.

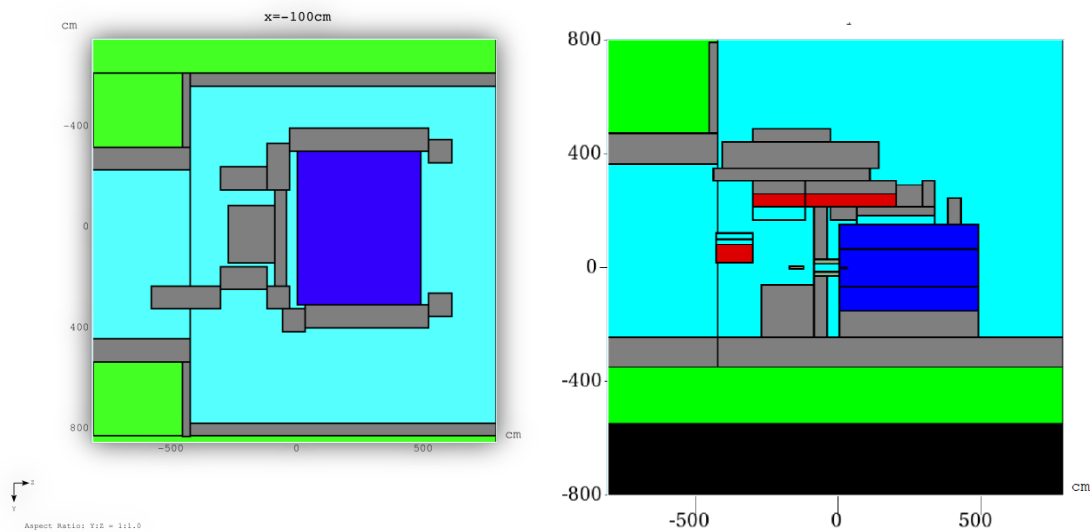

Figure 2: Left) MARS15 model, plan sectional view 100 cm below the beam axis. Grey is concrete, dark blue is iron, green is earth, red is steel, light blue is air. Right) MARS15 model, elevation sectional view on the beam axis.

SeaQuest will use two cryogenic liquid targets (hydrogen and deuterium) and three solid targets (carbon, calcium, and tungsten). The thickest target in terms of nuclear interaction lengths will be the 20 inch long (12% of $\lambda_I$) $LD_2$ target.

For this simulation, a proton beam intensity of 1.67E11 protons per second was assumed. This corresponds to 1E13 protons per minute and 3.2 kW of average beam power. The target simulated is 20 inches of $LD_2$. MARS15 simulates electromagnetic and hadronic showers by tracking individual particles, and includes processes such as gamma emission following neutron capture. The low energy cutoff for neutrons used in this simulation was 1 meV (milli-electron volt).

Essentially the only charged particles that escape FMAG are muons, which stay below grade. Most of the prompt dose outside the shielded region is due to neutrons or to gammas from neutron capture. Figure 3 shows plan and elevation views of the total prompt dose. Figure 4 shows the prompt dose due to neutrons, and Figure 5 shows the prompt dose due to muons.

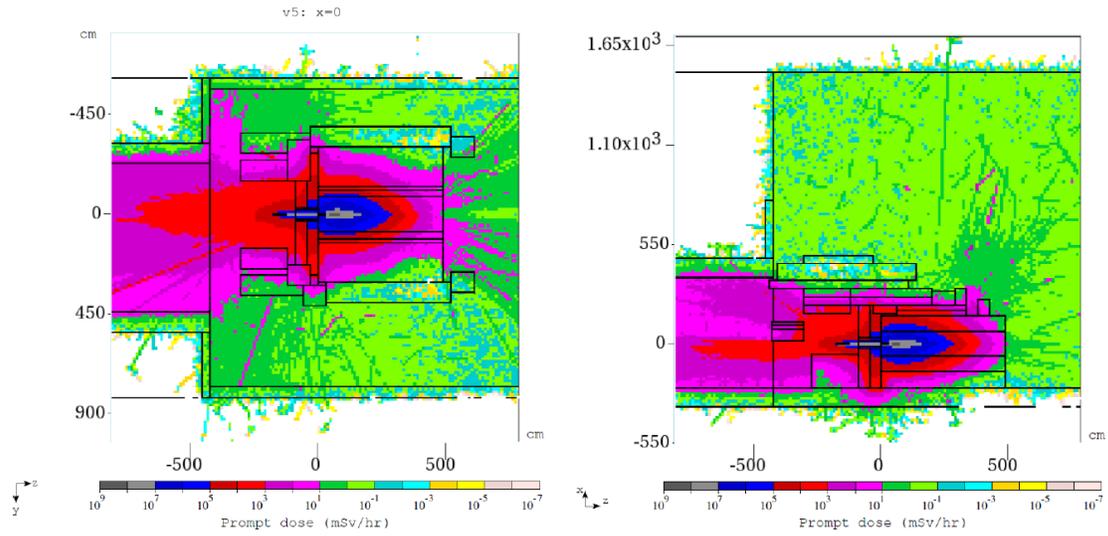

**Figure 3: Plan view (left) and elevation view (right) of the total prompt dose rate in milliSieverts per hour. The dose rate scale in these plots runs from $10^9$ mSv/hr to $10^{-7}$ mSv/hr ($10^{11}$ mrem/hr to $10^{-5}$ mrem/hr)**

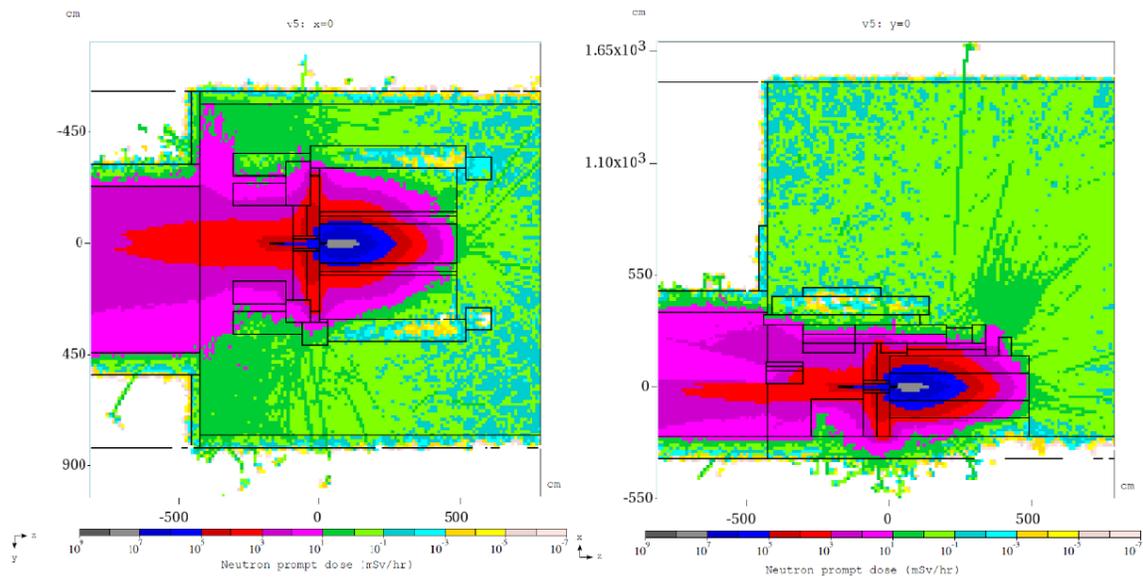

**Figure 4: Plan view (left) and elevation view (right) of the prompt dose rate due to neutrons (mSv/hr)**

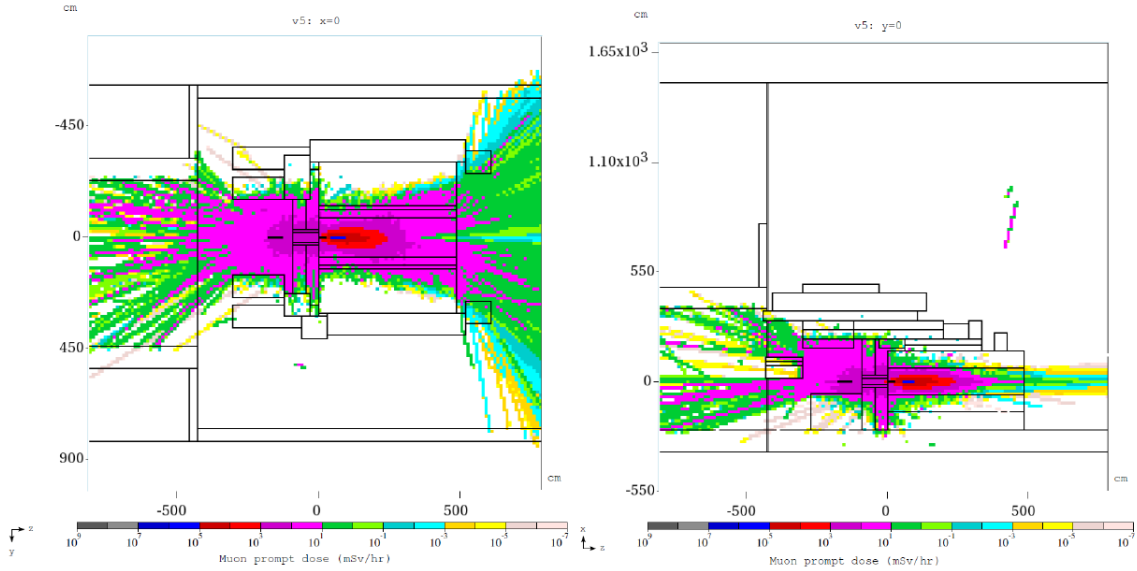

**Figure 5: Plan view (left) and elevation view (right) of the prompt dose rate due to muons (mSv/hr)**

Four points of interest were identified by the Neutrino Muon Shielding Assessment. We have estimated the dose rates at these points from the plots shown below. For most of these points there is additional shielding outside the volume simulated by MARS, and the dose rate including that shielding are given in Table 1. Additional shielding measurements for Points A and B were done using FESS Radiation Safety Drawings 9-8-6-12 C-4 and Point C used C-14.

| Point | Location | Dose Rate (mSv/hr) | Dose Rate (mrem/hr) | Additional shielding | Dose Rate after additional shielding (mrem/hr) | FRCM (mrem/hr) |
|---|---|---|---|---|---|---|
| A | Inside NM4 at upstream end of concrete cave - beam left | 1 | 100 | 5 feet of dirt | 1.64 | 5 ≤ DR ≤ 100 |
| B | Inside NM4 looking North or downstream - Parking lot | 1 | 100 | 8.5 feet of dirt | 0.092 | .05 ≤ DR ≤ 0.25 |
| C | Inside NM4 adjacent to NM4 gas shed | 0.1 | 10 | 8.5 feet of dirt | 0.0092 | .05 ≤ DR ≤ 0.25 |
| D | Inside NM4 roof above the upstream end of the concrete cave | 0.1 | 10 | N/A | 10 | 5 ≤ DR ≤ 100 |

**Table 1: Total prompt dose at points of interest (beam intensity = 1E13/minute)**

Points A and B are located just inside the NM3 and NM4 enclosure, respectively, at an elevation level of 750 ft. This location is at z=-315 cm (near the upstream end of the concrete cave) along the beamline. The corresponding x-y dose map is given in Figure 6 (with points A and B indicated). As can be seen in the expanded sections of the figure, there is a contour line between dark green and magenta located near both point A and point B. We estimate the dose rate at both points to be the value of this contour, which is 1 mSv/hr.

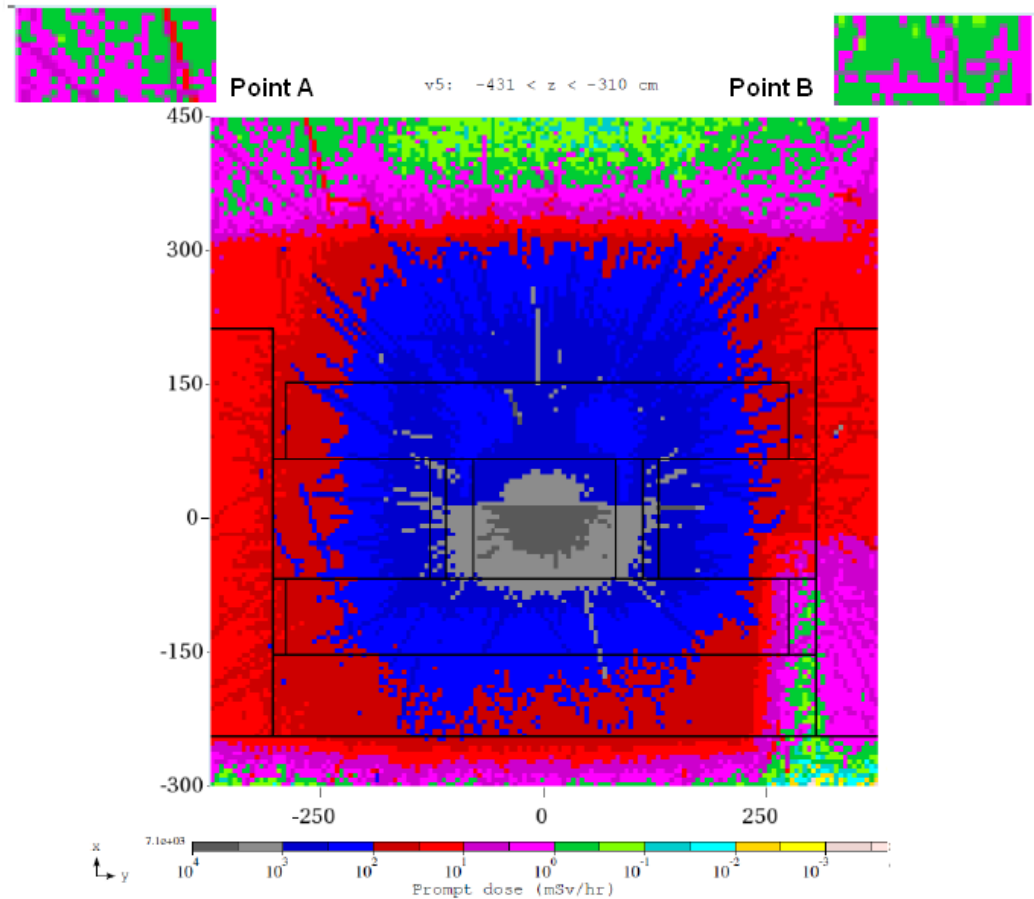

**Figure 6: Prompt dose through a transverse section at z=-315cm, including points A and B.**

Point C is located just inside NM4 at ground level, elevation 750 ft, at z=475 cm along the beamline. This location is relevant because it is adjacent to the NM4 gas shed. The corresponding x-y dose map is given in Figure 7. Looking at the contour lines we see a clear line of light red and magenta. This means a dose rate of 0.1 mSv/hr.

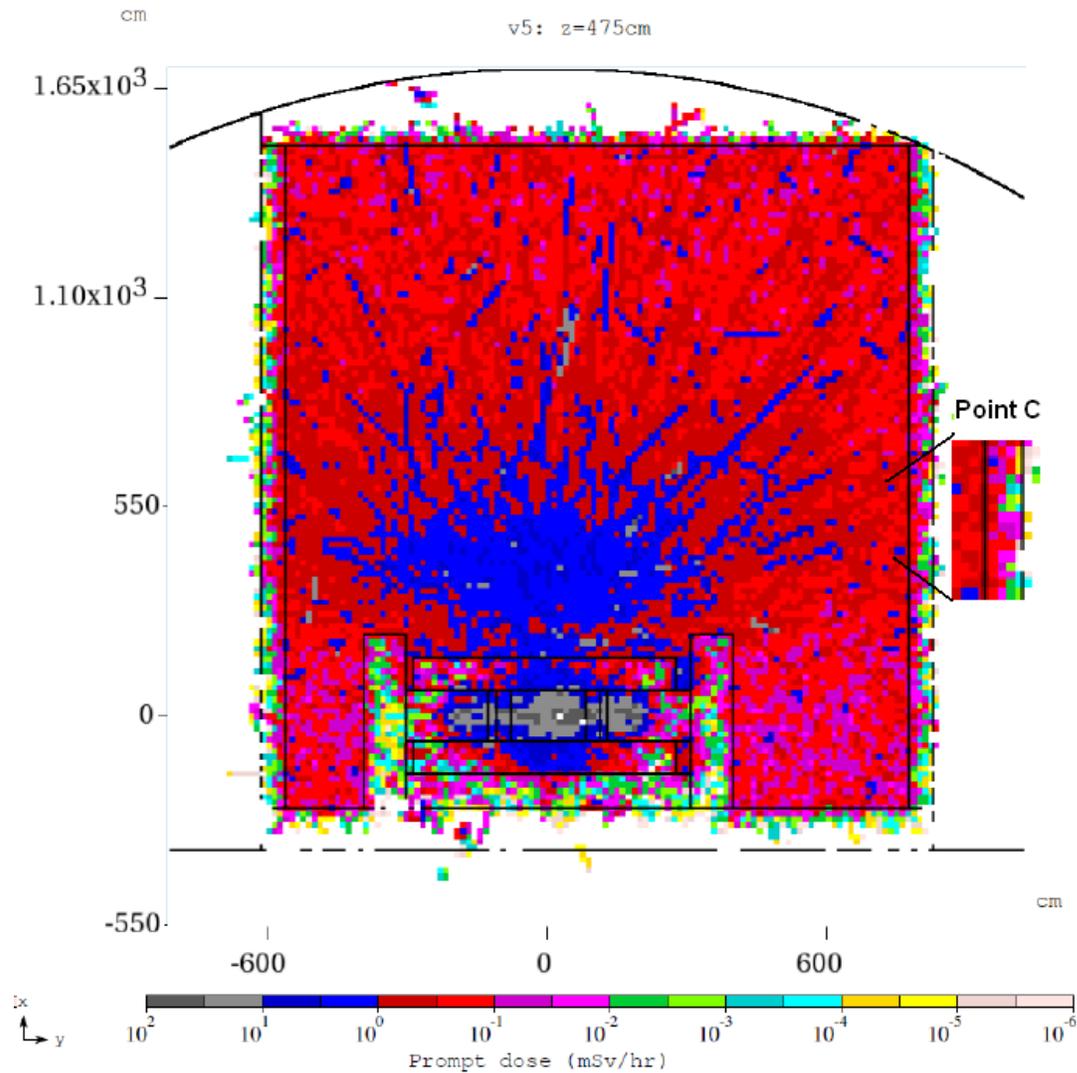

**Figure 7: Prompt dose through a transverse section at z=475 cm, including point C**

Point D is located at the roof of NM4 directly above the beam line where the MARS model includes steel and concrete on the roof of the target cave. The MARS model does not include a detailed representation of the metal roof of the NM4 building or the insulation and roofing material on top of the metal roof. As can be seen in the Figure 8, looking at the contour lines we see a clear line of light green and slate blue. We estimate the dose rate to be 0.1 mSv/hr.

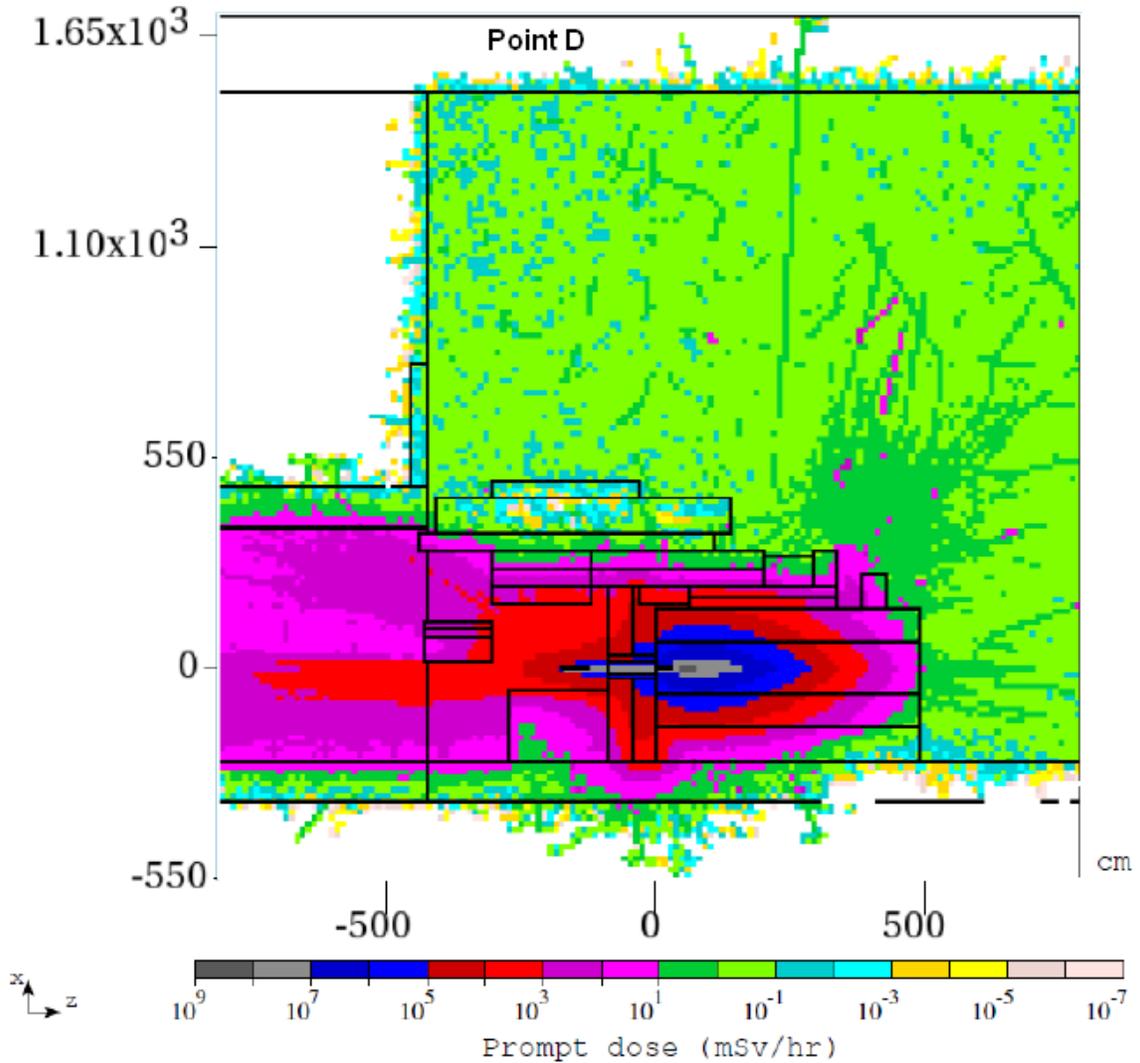

**Figure 8: Prompt dose through a longitudinal section on the beam line, including point D**

For the Ground Water and Surface Water calculations section of the Neutrino Muon Shielding Assessment, the maximum star density is required. As can be seen from the plan and elevation views in Figure 9, the maximum star density occurs under the upstream end of FMAG.

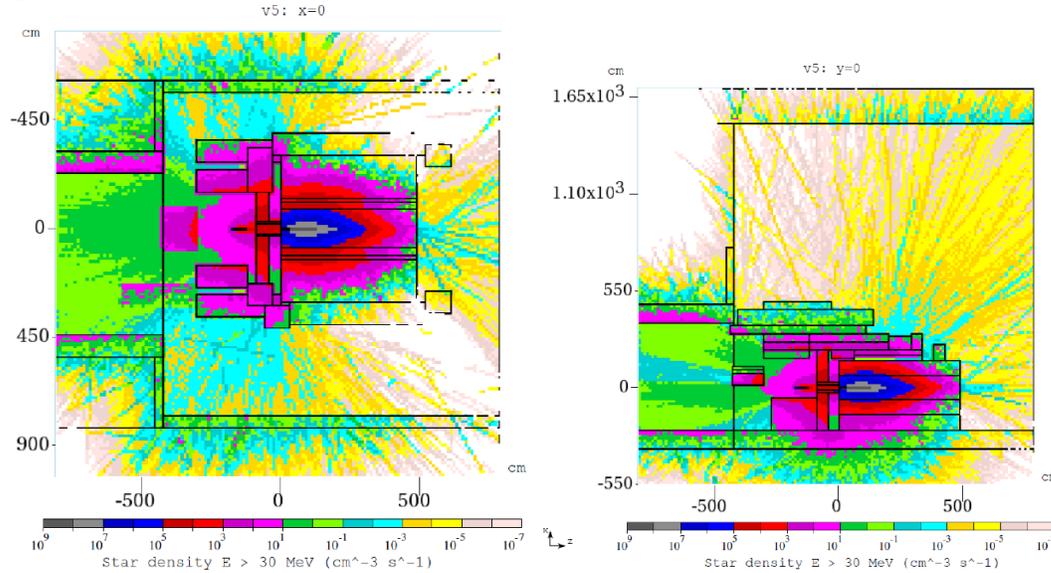

**Figure 9: Plan (left) and elevation (right) views of star density plots from which the maximum star density is found.**

Figure 10 shows the region of maximum star density in greater detail. The light purple shows a range of 10 stars per cubic centimeter to 100 stars per cubic centimeter. Conservatively taking the upper range which is 100 stars per cubic centimeter per second, when multiplying by the number of protons per seconds, 1.67E11, gives 5.99E-10 Stars per cubic centimeter per proton.

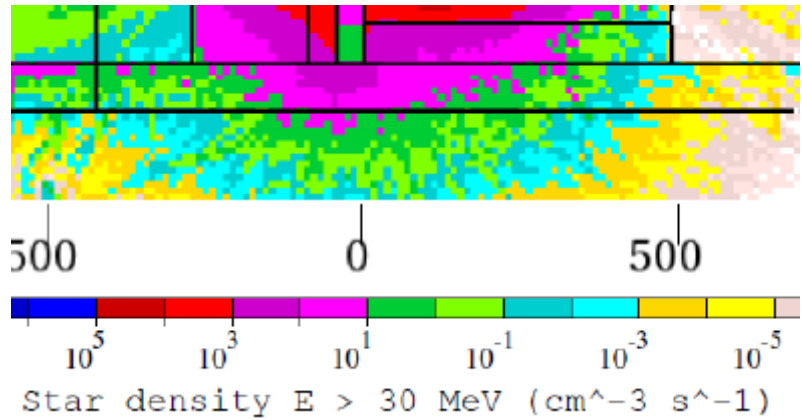

**Figure 10: Detail of the (elevation view) area of maximum star density in dirt under the concrete floor of NM4 near the upstream end of FMAG.**

Figure 11 shows the Residual Dose Rates at the target region of NM3 and NM4. The simulation was done for 30 days of running and 1 day of cool off. The maximum Dose Rate in the NM4 Experimental Hall and in the NM3 enclosure is seen to be the contour of light red to dark red. This contour corresponds to 0.1 mSv/hr. Not shown in the simulation is a fence at the upstream end of the target cave. This barrier isolates the target cave from the rest of the NM3 enclosure. The downstream end of the target cave is isolated from FMAG by 18 inches of concrete except in the area immediately surrounding the beam line. Inside this area the maximum dose rate can be seen to be dark red, corresponding to 1 mSv/hr.

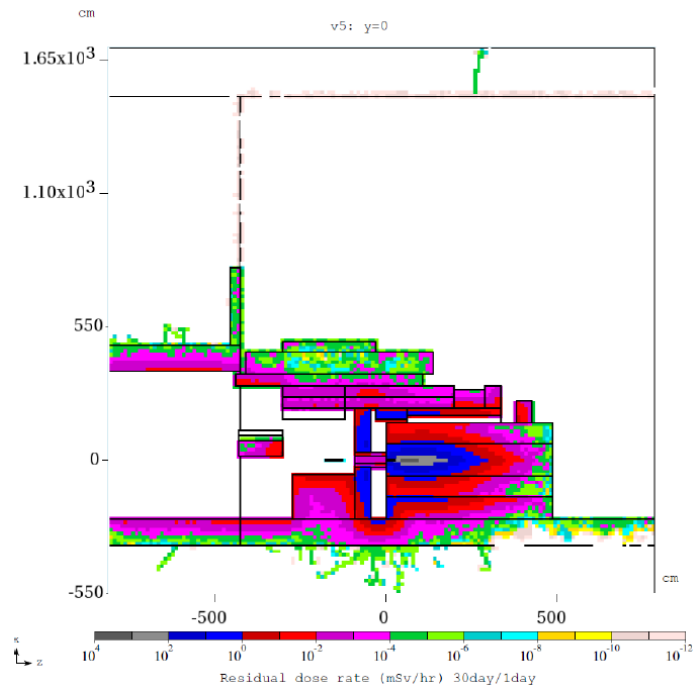

**Figure 11: Plan (left) and elevation (right) views Residual Dose Rate after 30 days running and one day of cool off. The scale for the Residual Dose Rate runs from $10^4$ mSv/hr to $10^{-12}$ mSv/hr ($10^6$ mrem/hr to $10^{-10}$ mrem/hr)**

Figure 12 shows the flux of hadrons with kinetic energy above 30 MeV from the simulation. The flux is used to calculate the air activation dose rates in "Air Activation Levels for the E906/SeaQuest Target Hall", Kamran Vaziri Jan 2011 and to calculate the need of a RAW system on FMag.

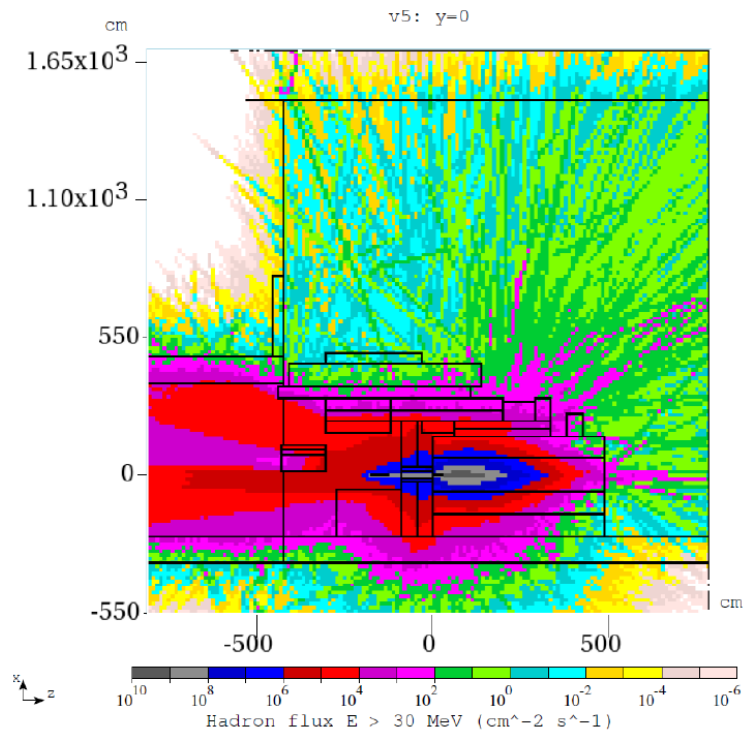

**Figure 12: Plan view of Hadron Flux of energy greater than 30 MeV**

References

N.V. Mokhov, "The Mars Code System User's Guide", Fermilab-FN-628 (1995);

O.E. Krivosheev, N.V. Mokhov, "MARS Code Status", Proc. Monte Carlo 2000 Conf., p. 943, Lisbon, October 23-26, 2000; Fermilab-Conf-00/181 (2000);

N.V. Mokhov, "Status of MARS Code", Fermilab-Conf-03/053 (2003);

N.V. Mokhov, K.K. Gudima, C.C. James et al, "Recent Enhancements to the MARS15 Code", Fermilab-Conf-04/053 (2004); http://www-ap.fnal.gov/MARS/.